\documentclass[a4paper]{JAC2003}

\usepackage{graphicx}
\usepackage{booktabs}

\begin{document}

\title{ A New Paradigm: Role of Electron-positron and Hadron Colliders  }

\author{Shou-hua Zhu
\\
$ ^1$ Institute of Theoretical Physics $\&$ State Key Laboratory of
Nuclear Physics and Technology,\\
Peking University, Beijing 100871, China \\
$ ^2$ Collaborative Innovation Center of Quantum Matter, Beijing, China \\
$ ^3$ Center for High Energy Physics, Peking University,
Beijing 100871, China
}

\maketitle

\begin{abstract}
   In 2012, a light scalar boson (denoted as H(125) in this paper) was discovered at the LHC.
We explore the possible correlation between the lightness of H(125) and the smallness of CP-violation
based on the Lee model, namely the spontaneous CP-violation two-Higgs-doublet-model.
It is a new way to understand why H(125) is light.
Based on this  we propose that it is the much heavier scalar bosons, instead of the H(125), which need to be understood. This opens
a new paradigm that one tries to understand the electro-weak symmetry breaking and CP violation. For the new paradigm, similar to many other
physics beyond the standard model, one need
both electron-positron and higher energy hadron collider, as well as the low energy experiments, in order to pin down the whole picture.
   \end{abstract}

\section{Introduction}

The organizers of HF2014 invited me to give an overview on physics, especially the physics beyond the standard model (BSM), which can be investigated at Higgs Factories (HF). Since
a new scalar (denoted as H(125) in this paper) was discovered in 2012, LHC is an obvious HF. For one hand, LHC can do much more in the future run, on the other hand, LHC precision is limited by its
hadronic environment. Next generation electron-positron collider and higher energy hadron collider are under extensive discussion. One predominant example is the CEPC (circular electron and  positron collider) with $\sqrt{s}= 240$ GeV or so, plus the possible update to super proton-proton collider (SPPC) with $\sqrt{s}=50-100$ TeV or higher. It is quite natural to expect
that CEPC can reach much higher precision that those of LHC, and SPPC can detect the much higher BSM scale than that of the LHC.

In principle, the whole BSM picture can usually be revealed via the combination efforts of LHC, CPEC/ILC, SPPC, FCC and other high energy colliders, as well as the low energy experiments which have certain unique opportunity, for example, CP violation and/or rare processes. There are numerous BSM, how to give the audience a relative global, objective and
persuasive picture is a challenging task. In the end we decide to firstly
give a brief overview on BSM motivations, then discuss a possible new paradigm as an example, which need high energy electron-positron and hadron colliders
to pin down the whole picture.

\section{Motivation for BSM}
 The discovery of H(125) is revolutionary. For the fist time in the history of particle physics, we have a complete theory to describe the electro-weak and strong interactions.
 If the H(125) is really the SM one, as many people believe, SM can be applicable to a very high scale, much higher than the weak scale. However we also have many reasons that BSM should exist.   In this section, the motivations for BSM are categorized into 4 classes.

 \subsection{Motivation(I): Test new types of interactions}

In the SM, there are 3 types of interactions:
\begin{itemize}
\item Gauge interaction
\item Yukwa interaction
\item Higgs self interaction
\end{itemize}

Gauge interaction, which can describe the strong and electro-weak interactions excellently, is well-tested for most cases.  The
Yukawa interaction and Higgs self-interaction are new types of interactions, and which need to be checked in HF.
In the SM Yukawa interaction is the origin of fermion mass, and induce the flavor changing processes. BSM can easily affect
Yukawa interaction. For the Higgs self interaction in the SM, once Higgs boson mass is fixed, the triple $h^3$ and quartic $h^4$ interactions
are also fixed. One motivation to measure the Higgs self couplings is related with electro-weak phase transition. In order
to account for the matter dominant Universe, the Higgs self couplings are usually greatly altered. Another popular motivation is that
Higgs potential might be more complicated than that in the SM. Therefore measuring the Higgs self interaction is the way to reconstruct
the Higgs potential, though quite challenging.

 \subsection{Motivation (II): Account for astrophysical observations}

There are several astrophysical observations (dark matter (DM), baryon asymmetry, and inflation) which may be related to BSM at O(TeV).
There is not suitable DM candidate in the SM. In order to keep DM stable or pseudo-stable, one usually introduces new symmetry, and one popular example
is the supersymmetry (SUSY). SUSY can provide a natural SM candidate is thought as one of its successes.  In order to construct a complete theory,
one likely introduces more Higgs fields, for example in the supersymmetric model we have to introduce at least two Higgs doublets.

SM can't account for baryon asymmetry in the Universe, though the 3 Saharov conditions are fulfilled.  If one insists on the electro-weak baryon-genesis,
the Higgs couplings must be altered, which can be tested in HF. Another possible connection between baryon asymmetry and HF is the CP violation. New CP violation other than
CKM matrix may be revealed in HF.

 \subsection{Motivation (III): Theoretical ones}

There are too many parameters in SM which could be reduced if we know what beneath the Higgs field(s). The introduction of scalar field brings some shortcomings to the
theory, for example the naturalness issue. In order to make the theory natural, one has to introduce new strong interaction, new symmetry, or extra-dimension etc.
There are many other theoretical issues, for example: what is the origin of P and CP violation, why are there 3 generations etc.

Last but not least, we don't include the gravity in the theory. HF may shed light on the nature of gravity, reveal hints for further knowledge on the space-time and quantum theory.

\subsection{Motivation (IV): "Common sense"}

For the fundamental science, exploring the unknown and discovering the unexpected phenomena are the primary driving force for the progress. We call this as
"common sense" motivation. To pursue the higher energy and smaller distance is the obvious next goal.

As the second "common sense" motivation, H(125) is special. Since the discovery of H(125) we have this new physical object which need the experimentalists to measure its properties as precisely as possible and discover possible deviations from the SM predictions.

\section{Smallness of CP violation and the lightness of H(125)}

\subsection{A story}

In 2006, officials of both PRC and US signed a document which indicated that a joint workshop would be held in June 2007, in order to explore the possibility of
parity spontaneous breaking at electron-positron colliders. In January 2007, a mini-workshop was organized by Prof. K.T. Chao and Prof. Y.P. Kuang
in order to prepare for the workshop. I was invited to give a theoretical and experimental review on left-right symmetric
models. During that time, the popular parity restoration models is left-right symmetric models. My personal impression on this model is that
the scale is too high to electron-positron colliders, even for LHC, because the low energy measurements like $K_0-\bar K_0$ mixing and direct search
from Tevatron have put
very strict limits on right-handed gauge bosons. After the mini-workshop, I realized that other parity restoration approaches, like mirror models, might be tested
at electron-positron collider and LHC \cite{Li:2007bz}.
In fact the parity of weak interaction is maximally violated, while CP is a quite good approximate symmetry. This is one of reasons why the left-right symmetric models suffering
so severe constraints. During the June workshop, the honorary chair of which is T.D. Lee, my impression on this topic kept the same.

In July 2012, the new scalar H(125) was discovered by ATLAS and CMS of the LHC, which surprised many theorists including me. In the SM, the discovery is not strange. However in the BSM, the
scalar mass tends to be the BSM scale. We were thinking possible new mechanism to account for the new discovery. In the end of 2012, just before the celebration of 80's birthday
of Prof. Kuang, I wrote the paper on the possible connection between the spontaneous CP violation and the mass of the new scalar \cite{Zhu:2012yv,Hu:2013cda}. Some people concerned on the viability of
the model. After more than one year, after checking all important measurements, we found \cite{Mao:2014oya} the model does viable and can be tested in the future LHC and electron-positron
colliders. Meanwhile we began to think the possible implication for understanding electro-weak symmetry breaking and origin of CP violation.

\subsection{Smallness of  CP violation}

CP has been thought strict conserved after the discovery of P violation in 1957. However such situation did not last long, tiny CP violation was first discovered in neutral K-meson in 1964. Now all measured CP violated effects in neutral K- and B-meson, and charged B meson systems can
be successfully accounted for by
the CKM matrix, which is usually parameterized as the Wolfenstein formalism
\begin{eqnarray}
V_{\textrm{CKM}} &=& \left(\begin{array}{ccc}1-\lambda^2/2&\lambda&A\lambda^3(\rho-i\eta)\\
-\lambda&1-\lambda^2/2&A\lambda^2\\A\lambda^3(1-\rho-i\eta)&-A\lambda^2&1\end{array}\right) \nonumber \\
&& +\mathcal{O}(\lambda^4)
\end{eqnarray}
In order to measure magnitude of CP violation independent on basis, the Jarlskog invariant \cite{PDG}\cite{Jarskog} was introduced as
\begin{equation}
J=A^2\lambda^6\eta=(2.96^{+0.20}_{-0.16})\times10^{-5}.
\end{equation}
The smallness of $J$ means the smallness of CP-violation
in the SM. In order to understand the smallness of CP violation, one has to go beyond the SM.

\subsection{Lightness of H(125)}

BSM is well motivated as we have discussed in the last section. BSM scale is usually pushed to a much higher value than that of weak interaction, given
the great success of the SM. In such circumstance, the 125 GeV scalar boson is unnatural, i.e. the lightness of the new scalar must link to certain mechanism.
It should be emphasized that the issue of the lightness of H(125) differs in the SM and the BSM.
In the SM, the mass
of Higgs boson is only a free parameter, and $m_H= 125$ GeV simply implies that the interactions are in the weak regime.
However in some BSM models there exists a light scalar naturally. We enumerate several examples, (1) in the minimal
super-symmetric model (MSSM), the lightest Higgs boson mass should be less than 140GeV after including higher-order corrections; (2) in the little higgs models,
 a Higgs boson which is
treated as a pseudo-Nambu-Goldstone boson must be light due to classical global symmetry and it acquires mass
through quantum effects only; (3) the anomalous in scale invariance models
can also generate a light Higgs boson; (4) the lightness of Higgs boson can intimately connect with the spontaneous CP violation.
While the first three approaches base on the {\em conjectured } symmetry, the last one utilizes the {\em  observed} approximate CP symmetry.
Historically Lee proposed the spontaneous CP violation in 1973 \cite{Lee} as an alternative way to induce CP violation. For the fourth approach, Lee's idea is extended to account for the lightness of the observed Higgs boson.

\subsection{Lee Model}

In order to analyze the connection between spontaneous CP violation and the Higgs boson mass,
we begin with the description of Lee model \cite{Lee} assuming that in the whole lagrangian there are no explicit CP-violated terms,
which means all the CP-violated effects come from a complex vacuum.

For the Lee model, the interactions of scalar fields can be written as \cite{Lee}
\begin{equation}
\mathcal{L}=(D_{\mu}\phi_1)(D^{\mu}\phi_1)+(D_{\mu}\phi_2)(D^{\mu}\phi_2)-V(\phi_1,\phi_2).
\end{equation}
Here
\begin{equation}
\label{vev}
\phi_1=\left(\begin{array}{c}\phi_1^+\\ \frac{v_1+R_1+iI_1}{\sqrt{2}}\end{array}\right),\quad\quad
\phi_2=\left(\begin{array}{c}\phi_2^+\\ \frac{v_2e^{i\xi}+R_2+iI_2}{\sqrt{2}}\end{array}\right)
\end{equation}
are the two Higgs doublets. We can see the $\xi$ is the only source of CP violation. Here there are two vacuum expectation values $v_1, v_2$ and $v=\sqrt{v_1^2+v^2_2}=246\textrm{GeV}$ as usual. Defining $R(I)_{ij}$ as
the real(imaginary) part of $\phi_i^{\dag}\phi_j$, we can write a general potential as
\begin{eqnarray}
V&=&V_2+V_4\nonumber\\
&=&\mu_{1}^2R_{11}+\mu_{2}^2R_{22}\nonumber\\
&&+\lambda_1R_{11}^2+\lambda_2R_{11}R_{12}+\lambda_3R_{11}R_{22}\nonumber\\
\label{pot}&&+\lambda_4R_{12}^2+\lambda_5R_{12}R_{22}+\lambda_6R_{22}^2+\lambda_7I_{12}^2.
\end{eqnarray}

We can also write the general Yukawa couplings as
\begin{eqnarray}
\mathcal{L}_y &=& -\bar{Q}_{Li}(Y_{1d}\phi_1+Y_{2d}\phi_2)_{ij}D_{Rj} \nonumber \\
&& -\bar{Q}_{Li}(Y_{1u}\tilde{\phi}_1+Y_{2u}\tilde{\phi}_2)_{ij}U_{Rj},
\end{eqnarray}
in which $\tilde{\phi}_i=\textrm{i}\sigma_2\phi_i^*$ and all Yukawa couplings are real.

After rotating away Goldstone mode, we can expand the neutral Higgs boson mass matrix $\tilde{m}$ in series of $t_{\beta}(s_{\xi})$ as
\begin{equation}
\tilde{m}=\tilde{m}_0+(t_{\beta}s_{\xi})\tilde{m}_1+(t_{\beta}s_{\xi})^2\tilde{m}_2+\cdots
\end{equation}
We found
\begin{equation}
\mathop{\lim}_{t_{\beta}s_{\xi}\rightarrow0}\det(\tilde{m})=\det(\tilde{m}_0)=0
\end{equation}
which means a zero eigenvalue of $\tilde{m}_0$ thus there must be a light neutral scalar when $t_{\beta}s_{\xi}$ is small.
To the leading order of $t_{\beta}s_{\xi}$, for the lightest scalar $h$, we have
\begin{eqnarray}
\label{mh}
m^2_h&=&\frac{v^2t^2_{\beta}s^2_{\xi}}{2}\left(\frac{(\tilde{m}_1)^2_{12}}{(\tilde{m}_0)_{22}}
+\frac{(\tilde{m}_1)^2_{13}}{(\tilde{m}_0)_{33}}+(\tilde{m}_2)_{11}\right) \nonumber\\
h&=&I_2+t_{\beta}s_{\xi}\left(\frac{(\tilde{m}_1)_{12}}{(\tilde{m}_0)_{22}}(c_{\theta}R_1+s_{\theta}R_2)
 \right. \nonumber \\
 && \left. +\frac{(\tilde{m}_1)_{13}}{(\tilde{m}_0)_{33}}(c_{\theta}R_2-s_{\theta}R_1)-\frac{I_1}{t_{\xi}}\right),
\end{eqnarray}
where $\theta=(1/2)\tan^{-1}(2\lambda_2/(4\lambda_1-\lambda_4+\lambda_7))$.

From the Yukawa couplings we will get the mass matrixes for fermions as
\begin{eqnarray}
\label{m1}(M_U)_{ij}=\frac{v}{\sqrt{2}}(Y_{1u}c_{\beta}+Y_{2u}s_{\beta}e^{-i\xi})_{ij},\\
\label{m2}(M_D)_{ij}=\frac{v}{\sqrt{2}}(Y_{1d}c_{\beta}+Y_{2d}s_{\beta}e^{i\xi})_{ij}.
\end{eqnarray}
We can always perform the diagonalization for $M_{U(D)}$ with matrixes $U(D)_L$ and $U(D)_R$ as
\begin{equation}
U_LM_UU_R^{\dag}=diag\left(m_u,m_c,m_t\right),
\end{equation}
\begin{equation}
D_LM_DD_R^{\dag}=diag\left(m_d, m_s, m_b \right).
\end{equation}
And $V_{\textrm{CKM}}=U_LD_L^{\dag}$ is the CKM matrix.

We choose all the nine free parameters as nine observables in Higgs sector: masses of four scalars $m_h,m_2,m_3$ and $m_{H^{\pm}}$;
vacuum expected values $v_1,v_2,\xi$ and two mixing angles for neutral bosons. We choose them $c_1$ and $c_2$ defined as
\begin{equation}
\mathcal{L}_{h_iVV}=c_ih_i\left(\frac{2m^2_W}{v}W^+_{\mu}W^{\mu-}+\frac{m^2_Z}{v}Z_{\mu}Z^{\mu}\right)
\end{equation}
which just means the $h_iVV$ vertex strength comparing with that in SM. There is a sum rule
$c_1^2+c_2^2+c_3^2=1$ due to spontaneous electro-weak symmetry broken, thus only two of the $c_i$ are free.

In the scalar sector, for non-degenerate neutral Higgs
bosons, a quantity $K=c_1c_2c_3$ measures the CP violation effects \cite{2HDM}\cite{K},
while in Yukawa sector, the Jarlskog invariant $J$ \cite{Jarskog} measures that. In this scenario,
to the leading order of $t_{\beta}s_{\xi}$, we have
\begin{equation}
\label{K}K=c_1c_2c_3=-s_{\theta}c_{\theta}(1+\eta_1)t_{\beta}s_{\xi}\propto t_{\beta}s_{\xi}
\end{equation}

In order to calculate J, we define matrix $\hat{C}$ as
\begin{equation}
\hat{C}\equiv \left[M_UM_U^{\dag},M_DM_D^{\dag}\right].
\end{equation}
We can always choose a basis in which the diagonal elements of $\hat{C}$ are zero. Thus
\begin{eqnarray}
\hat{C} &=& \left(\begin{array}{ccc}0&C_3&-C_2\\-C_3&0&C_1\\C_2&-C_1&0\end{array}\right)+
i\left(\begin{array}{ccc}0&C_3^*&C_2^*\\C_3^*&0&C_1^*\\C_2^*&C_1^*&0\end{array}\right) \nonumber \\
&& =
\left(\textrm{Re}\hat{C}+i\textrm{Im}\hat{C}\right)
\end{eqnarray}
It can be proved that \cite{Mao:2014oya}
\begin{equation}
\label{J}J=\frac{\prod C_i\sum(C_i^*/C_i)}{\prod\left(m^2_{U_i}-m^2_{U_j}\right)\prod\left(m^2_{D_i}-m^2_{D_j}\right)}\propto t_{\beta}s_{\xi}.
\end{equation}
According to the equations (\ref{K}), (\ref{J}), and (\ref{mh}), we propose that the lightness of the Higgs
boson and the smallness of CP-violation effects could be correlated through small $t_{\beta}s_{\xi}$ since both
the Higgs mass $m_h$ and the quantities $K$ and $J$ to measure CP-violation effects are proportional to
$t_{\beta}s_{\xi}$ at the small $t_{\beta}s_{\xi}$ limit.

\section{A new paradigm}

\begin{figure}[!tbh]
  \centering
   \includegraphics[width=\linewidth]{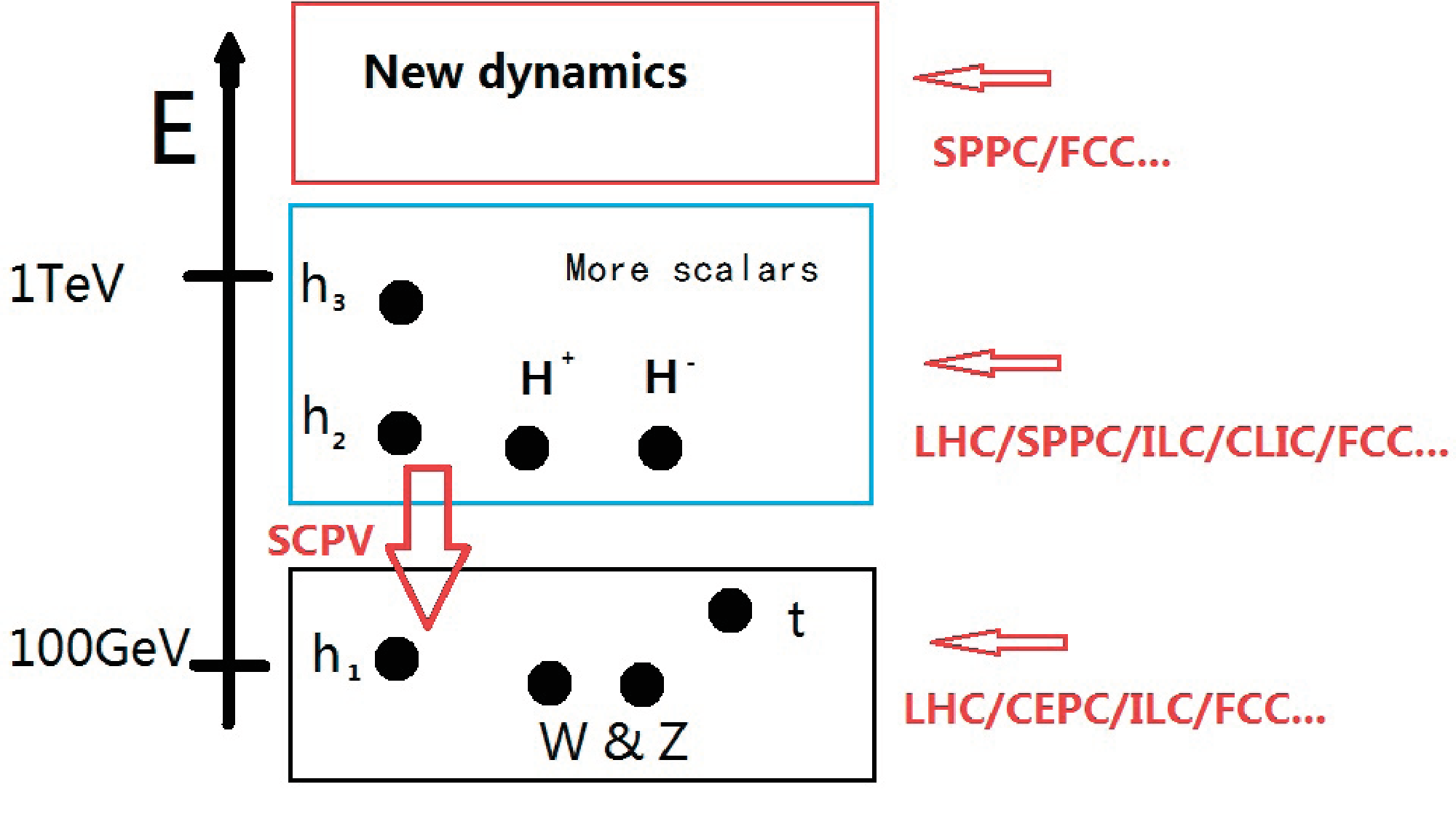}
      \caption[]{ Schematic diagram for a new paradigm.}
         \label{anp}
  \end{figure}

In the last section, the connection between small CP violation and H(125) is explored in the simplest spontaneous CP violation Lee model.
Based on the above discussion, a new paradigm emerges. In the past studies, one tried to account for the lightness of the H(125). In the new
paradigm, the mass of H(125) is due to the  approximate CP symmetry and the extra scalar bosons can be much heavier, e.g. at O(TeV) even in
the strong-coupled regime. This
opens a new approach to understand the electro-weak symmetry breaking and the origin of CP violation. The schematic diagram is shown in
Fig. \ref{anp}. For the optimistic case (from the point view of experimentalists), the extra new scalars can be not so heavy.
For this case, LHC has the good opportunity to discover them.

The generic feature of this new paradigm is summarized
\begin{itemize}
\item H(125) is not SM-like, instead H(125) is the CP mixing state
\item There are other heavy neutral and charged Higgs bosons
\item There are usually extra CP violation in scalar sector, besides CKM matrix, though they arise both from the complex vacuum
\item The scale of the new mechanism for the complicated scalar sectors is higher.
\end{itemize}

\section{Phenomenology: Role of electron-positron and hadaron colliders}

Based on studies of Ref. \cite{Mao:2014oya}, the Lee model is still viable confronting
the high and low energy experiments. The next natural question is how to confirm/exclude this model at future facilities.

We treat the H(125) as the lightest neutral Higgs boson. If Lee model is true, the extra neutral and charged Higgs bosons should be discovered at high energy colliders.
As the general rules, the lighter the extra Higgs bosons, the easier they can be produced.
In order to confirm the Lee model, another possible signal can be the FCNC decay of the neural Higgs bosons which are unobservable small in the SM.
Furthermore the CP properties of the Higgs boson are essential measurements, though it is a very challenging task.

As we have pointed out  \cite{Mao:2014oya} that there is no SM limit in this scenario, thus it is always testable at the future colliders,
such as LHC with $\sqrt{s}=14\textrm{TeV}$, CEPC,
ILC, or TLEP with $\sqrt{s}=(240\sim250)\textrm{TeV}$, even before the discovery of other neutral Higgs bosons and charged Higgs
boson.
The coupling between H(125) and
other particle (especially $W^{\pm}$ and $Z^0$)
are usually suppressed by the factor of $\mathcal{O}(t_{\beta}s_{\xi})$.
In the $b\bar{b}$ decay channel or any VBF, VH associated production
channel, a significant suppression can be the first sign of this scenario. On the contrary if the signals become even more SM-like,
this scenario will be disfavored.

For future LHC with $\sqrt{s}=14\textrm{TeV}$, the signal strengths will be measured with an uncertainty of about
$10\%$ at the luminosity $300\textrm{fb}^{-1}$ \cite{cms14}\cite{atl14}. These measurements can be used to constrain $c_V$ ( which is $c_1$ in the last section).
 A Higgs boson with $c_V \geq (0.6\sim0.7)$ is hardly to be pseudoscalar dominant thus if
$c_V \leq (0.6\sim0.7)$ is excluded, we can say this scenario is excluded. So we can test this scenario by fitting the
signal strengths. We list the estimating results in Tab. \ref{excl}.
\begin{table}[h]
\centering
\caption{Abilities to test the scenario  at $\sqrt{s}=14\textrm{TeV}$ LHC. Lower limit for the allowed $c_V$ at
$2\sigma$ and $3\sigma$ level are listed in the tables. For the up/down tables we assume all signal strengths are consistent
with SM at $1\sigma$/$2\sigma$ level respectively.}\label{excl}
\begin{tabular}{|c|c|c|}
\hline
Excluded level&$2\sigma$&$3\sigma$\\
\hline
$300\textrm{fb}^{-1}$&$0.62$&$0.55$\\
\hline
$3000\textrm{fb}^{-1}$&$0.77$&$0.72$\\
\hline
\end{tabular}
\quad\quad
\begin{tabular}{|c|c|c|}
\hline
Excluded level&$2\sigma$&$3\sigma$\\
\hline
$300\textrm{fb}^{-1}$&$0.53$&$0.45$\\
\hline
$3000\textrm{fb}^{-1}$&$0.7$&$0.65$\\
\hline
\end{tabular}
\end{table}
It should be noted that the true ability to test this scenario depends strongly on the
real signal strengths from future experiments.

At a HF with the $e^+e^-$ initial state at $\sqrt{s}=(240\sim250)\textrm{GeV}$, the dominant production process
for a Higgs boson is associated with a $Z^0$ boson. Another important production process is through VBF. In this scenario
it is suppressed by a factor $c_1^2$ thus this scenario can be excluded if the total cross section favors SM. For the total
cross section, a measurement with $\mathcal{O}(10\%)$ uncertainty is accurate enough to distinguish the scenario we discussed
in this paper and SM at $3\sigma$ or even $5\sigma$ significance.
Such accuracy can be easily achieved at CEPC/ILC/TLEP.

For this new paradigm, the role of the higher energy hadron collider (like SPPC) is to probe the new dynamics which is responsible for
the electro-weak symmetry breaking, origin of CP and/or P violation etc.

\section{Conclusions and discussions}

In this paper, we explored the correlation between the lightness of Higgs boson and the smallness of CP-violation
based on the Lee model, namely the spontaneous CP-violation two-Higgs-doublet-model.
In this model, the mass of the lightest Higgs boson $m_h$ as well as the quantities $K$
and $J$ are $\propto t_{\beta}s_{\xi}$
in the limit $t_{\beta}s_{\xi}\rightarrow0$, namely the CP conservation limit. Here $K$
and $J$ are the measures for CP-violation effects in scalar and Yukawa sectors respectively.
It is a new way to understand why the Higgs boson discovered at the LHC is light.
Based on this possible connection, we proposed that H(125) is due to the approximate CP symmetry and there are extra heavier scalars. This opens
a new paradigm that one tries to understand the electro-weak symmetry breaking and CP violation. For this new paradigm, one needs
both electron-positron and higher energy hadron collider, as well as the low energy experiments, to pin down the whole picture. We emphasize that
this final conclusion is also applicable to many other BSM as well, and which is one of the most important motivations to pursue the next generation colliders.

\subsection*{Acknowledgment}

 This work was supported in part by the Natural Science Foundation
 of China (Nos. 11135003 and 11375014).

\end{document}